\documentclass[12pt]{article}

\usepackage{standard}

\newcommand{\st}{\mathcal{S}}
\newcommand{\cst}{\text{\rm co-}{\mathcal{S}}}
\newcommand{\subst}{{\downarrow\!\st}}
\newcommand{\csubst}{{\downarrow\!\cst}}

\def\yes{\text{yes}}
\def\no{\text{no}}

\begin{document}

\title{Toward a General Theory of Quantum Games}

\author{
  Gus Gutoski \hspace{2cm} John Watrous\\[2mm]
  Institute for Quantum Computing\\
  and School of Computer Science\\
  University of Waterloo\\
  Waterloo, Ontario, Canada
}

\maketitle

\begin{abstract}
  We study properties of {\it quantum strategies}, which are
  complete specifications of a given party's actions in any
  multiple-round interaction involving the exchange of quantum
  information with one or more other parties.
  In particular, we focus on a representation of quantum strategies that
  generalizes the Choi-Jamio{\l}kowski representation of quantum
  %operations, with respect to which each strategy is described by a single
  operations.  This new representation associates with each strategy a
  positive semidefinite operator acting only on the tensor product of
  its input and output spaces.
  Various facts about such representations are established,
  and two applications are discussed: the first is a new and
  conceptually simple proof of Kitaev's lower bound for strong
  coin-flipping, and the second is a proof of the exact
  characterization $\class{QRG} = \class{EXP}$ of the class of
  problems having quantum refereed games.
\end{abstract}

%=============================================================================%

\section{Introduction}

\noindent
The theory of games provides a general structure within which both
cooperation and competition among independent entities may be modeled,
and provides powerful tools for analyzing these models.
Applications of this theory have fundamental importance in many
areas of science.

This paper considers games in which the players may exchange and process
quantum information.
We focus on competitive games, and within this context the types of
games we consider are very general.
For instance, they allow multiple rounds of interaction among the
players involved, and place no restrictions on
players' strategies beyond those imposed by the theory of quantum information.

While classical games can be viewed as a special case of quantum games, it
is important to stress that there are fundamental differences between
general quantum games and classical games.
For example, the two most standard representations of classical
games, namely the {\it normal form} and {\it extensive form}
representations, are not directly applicable to general quantum games.
This is due to the nature of quantum information, which admits a continuum
of pure (meaning extremal) strategies, imposes bounds on players' knowledge
due to the uncertainty principle, and precludes the representation of
general computational processes as trees.
In light of such issues, it is necessary to give special consideration
to the incorporation of quantum information into the theory of games.

A general theory of quantum games has the potential to be useful in many
situations that arise in quantum cryptography, computational complexity,
communication complexity, and distributed computation.
This potential is the primary motivation for the work presented in this paper,
which we view as a first step in the development of a general theory of quantum
games.
The following facts are among those proved in this paper:

\begin{itemize}
\item[$\bullet$]
  Every multiple round quantum strategy can be faithfully represented by a
  single positive semidefinite operator acting only on the
  tensor product of the input and output spaces of the given player.
  This representation is a generalization of the Choi-Jamio{\l}kowski
  representation of super-operators.
  The set of all operators that arise in this way is precisely characterized
  %by the positive semidefinite constraint along with a simple collection of
  %linear constraints.
  by the set of positive semidefinite operators that satisfy a simple 
  collection of linear constraints.

\item[$\bullet$]
  %Multiple round quantum strategies including measurements are represented
  %in a similar way, with one positive semidefinite operator corresponding to
  %each measurement result.
  %The probability of a given pair of measurement outcomes for two interacting
  %strategies is given by the inner product of their representations.
  If a multiple round quantum strategy calls for one or more measurements then
  its representation consists of one operator for each of the possible
  measurement outcomes.
  The probability of any given pair of measurement outcomes for two interacting
  strategies is given by the inner product of their associated operators.

\item[$\bullet$]
  The maximum probability with which a given strategy can be forced to output
  a particular result is
  %of course given by an optimization over an opposing
  %player's strategies.
  %A more surprising alternate description of this maximum probability
  %is that it is
  the {\it minimum} value of $p$ for which the positive semidefinite
  operator corresponding to the given measurement result is bounded
  above (with respect to the L{\"o}wner partial order) by the
  representation of a valid strategy multiplied by~$p$.
\end{itemize}

\noindent
We give the following applications of these facts:

\begin{itemize}
\item[$\bullet$]
  A new and conceptually simple proof of Kitaev's bound for strong
  coin-flipping, which states that every quantum strong coin-flipping
  protocol allows a bias of at least $1/\sqrt{2} - 1/2$.

\item[$\bullet$]
  The exact characterization $\class{QRG} = \class{EXP}$ of the
  class of problems having quantum refereed games
  (i.e., quantum interactive proof systems with two competing provers).
  This establishes that quantum and classical refereed games are equivalent
  in terms of expressive power: $\class{QRG} = \class{RG}$.
\end{itemize}

\subsubsection*{Relation to previous work}

It is appropriate for us to comment on the relationship between the present
paper and a fairly large collection of papers written on a topic that has been
called {\it quantum game theory}.
Meyer's {\it PQ Penny Flip} game \cite{Meyer99} is a well-known example
of a game in the category these papers consider.
The work of Eisert, {\it et~al.}~\cite{EisertW+99} is also commonly cited in
this area.
Some controversy exists over the interpretations drawn in some quantum
these papers---see, for instance, Refs.~\cite{BenjaminH01-comment,EnkP02}.

A key difference between our work and previous work on quantum game
theory is that our focus is on multiple-round interactions.
Understanding the actions available to players that have quantum
memory is therefore critical to our work, and to our knowledge has not
been previously considered in the context of quantum game theory.

A second major difference is that, in most of the previous quantum
game theory papers we are aware of, the focus is on rather specific
examples of classical games and on identifying differences that arise
when so-called quantum variants of these games are considered.
As a possible consequence, it may arguably be said that none of the results
proved in these papers has had sufficient generality to be applicable to any
other studies in quantum information.
In contrast, our interest is not on specific examples of games, but rather on
the development of a general theory that holds for all games.
It remains to be seen to what extent our work will be applied, but the
applications that we provide suggest that it may have interesting uses in
other areas of quantum information and computation.

A different context in which games arise in quantum information
theory is that of {\it nonlocal games} \cite{CleveH+04}, which include
{\it pseudo-telepathy games} \cite{BrassardB+05} as a special case.
These are cooperative games of incomplete information that model situations
that arise in the study of multiple-prover interactive proof systems, and
provide a framework for studying Bell Inequalities and the notion of
nonlocality that arises in quantum physics.
While such games can be described within the general setting we
consider, we have not yet found an application of the methods of the
present paper to this type of game.
Possibly there is some potential for further development of our
work to shed light on some of the difficult questions in this area.

%=============================================================================%

\section{Preliminaries}
\label{sec:prelim}

\noindent
This section gives a brief overview of various quantum
information-theoretic notions that will be needed for the remainder of
the paper.
We assume the reader has familiarity with quantum information theory,
and intend only that this overview will serve to establish our
notation and highlight the main concepts that we will need.
Readers not familiar with quantum information are referred to the
books of Nielsen and Chuang \cite{NielsenC00} and Kitaev, Shen and
Vyalyi \cite{KitaevS+02}.

When we speak of the vector space associated with a given quantum
system, we are referring to some complex Euclidean space (by which we
mean a finite-dimensional inner product space over the complex numbers).
Such spaces will be denoted by capital script letters such as $\X$, $\Y$,
and $\Z$.
We always assume that an orthonormal {\it standard basis} of any such space
has been chosen, and with respect to this basis elements of these spaces
are associated with column vectors, and linear mappings from one space to
another are associated with matrices in the usual way.
We will often be concerned with finite sequences $\X_1,\,\X_2,\,\ldots,\,\X_n$
of complex Euclidean spaces.
We then define
\[
\X_{i\ldots j} = \X_i\otimes\cdots\otimes\X_j
\]
for nonnegative integers $i,j\leq n$, and define $\X_{i\ldots j}=\complex$ for
$i>j$.

It is convenient to define various sets of linear mappings between
given complex Euclidean spaces $\X$ and~$\Y$ as follows.
Let $\lin{\X,\Y}$ denote the space of all linear mappings
(or {\it operators}) from $\X$ to $\Y$, and write $\lin{\X}$ as
shorthand for $\lin{\X,\X}$.
We write $\herm{\X}$ to denote the set of Hermitian operators acting
on $\X$, $\pos{\X}$ to denote the set of all positive semidefinite
operators acting on $\X$, and $\density{\X}$ to denote the set of all
density operators on $\X$ (meaning positive semidefinite operators
having trace equal to 1.)
An operator $A\in\lin{\X,\Y}$ is a {\it linear isometry} if
$A^{\ast} A = I_{\X}$.
The existence of a linear isometry in $\lin{\X,\Y}$ of course requires that
$\dim(\X)\leq\dim(\Y)$, and if $\dim(\X)=\dim(\Y)$ then any linear isometry
$A\in\lin{\X,\Y}$ is unitary.
We let $\unitary{\X,\Y}$ denote the set of all linear isometries from
$\X$ to $\Y$.
The operator $I_{\X}\in\lin{\X}$ denotes the identity operator
on~$\X$.
Transposition of operators is always taken with respect to standard bases.

The Hilbert-Schmidt inner product on $\lin{\X}$ is defined by
\[
\ip{A}{B} = \tr(A^{\ast}B)
\]
for all $A,B \in \lin{\X}$.

For given operators $A,B\in\herm{\X}$, the notation $A\leq B$ means
that $B-A \in\pos{\X}$.
This relation is sometimes called the {\it L{\"o}wner partial order}
on $\herm{\X}$.

When we refer to {\it measurements}, we mean POVM-type measurements.
Formally, a measurement on a complex Euclidean space $\X$ is
described by a collection of positive semidefinite operators
\[
\left\{P_{a}\,:\,a\in\Sigma\right\}\subset\pos{X}
\]
satisfying the constraint
\[
\sum_{a\in\Sigma}P_a = I_{\X}.
\]
Here $\Sigma$ is a finite, non-empty set of {\it measurement outcomes}.
If a state represented by the density operator $\rho$ is measured with
respect to such a measurement, each outcome $a\in\Sigma$ results with
probability $\ip{P_a}{\rho} = \tr(P_a\rho)$.

A {\it super-operator} is a linear mapping of the form
\[
\Phi: \lin{\X} \rightarrow \lin{\Y},
\]
where $\X$ and $\Y$ are complex Euclidean spaces.
A super-operator of this form is said to be {\it positive} if
$\Phi(X)\in\pos{\Y}$ for every choice of $X\in\pos{\X}$, and is
{\it completely positive} if $\Phi\otimes I_{\lin{\Z}}$ is positive
for every choice of a complex Euclidean space $\Z$.
The super-operator $\Phi$ is said to be {\it admissible} if
it is completely positive and preserves trace:
$\tr(\Phi(X)) = \tr(X)$ for all $X\in\lin{\X}$.
Admissible super-operators represent discrete-time changes in quantum
systems that can, in an idealized sense, be physically realized.

The Choi-Jamio{\l}kowski representation \cite{Jamiolkowski72, Choi75}
of super-operators is as follows.
Suppose that $\Phi: \lin{\X} \rightarrow \lin{\Y}$ be a given super-operator
and let $\{\ket{1},\ldots,\ket{N}\}$ be the standard basis of $\X$.
Then the {\it Choi-Jamio{\l}kowski representation} of $\Phi$ is the
operator
\[
J(\Phi) = \sum_{1\leq i,j\leq N} \Phi(\ket{i}\bra{j}) \otimes \ket{i}
\bra{j} \in\lin{\Y\otimes\X}.
\]
It holds that $\Phi$ is completely positive if and only if $J(\Phi)$ is 
positive semidefinite, and that $\Phi$ is trace-preserving if and only if
$\tr_{\Y}(J(\Phi)) = I_{\X}$.

For two complex Euclidean spaces $\X$ and $\Y$, we define a linear mapping
\[
\op{vec}:\lin{\X,\Y} \rightarrow \Y\otimes\X
\]
by extending by linearity the action $\ket{i}\bra{j}\mapsto\ket{i}\ket{j}$
on standard basis states.
We make extensive use of this mapping in some of our proofs, as it is very
convenient in a variety of situations.
Let us now state some identities involving the $\op{vec}$ mapping, each of
which can be verified by a straightforward calculation.
\begin{prop}
\label{prop:identities}
The following hold:
\begin{mylist}{5mm}
\item[1.]
For any choice of $A$, $B$, and $X$ for which the product $A X B^{\t}$
makes sense we have
\[
(A\otimes B)\op{vec}(X) = \op{vec}(AXB^{\t}).
\]

\item[2.]
\label{item:trace}
For any choice of $A,B\in\lin{\X,\Y}$ we have
\begin{align*}
\tr_{\X}(\op{vec}(A)\op{vec}(B)^{\ast}) & = AB^{\ast}, \\
\tr_{\Y}(\op{vec}(A)\op{vec}(B)^{\ast}) & = \left( B^{\ast}
  A\right)^{\t}.
\end{align*}

\item[3.]
For any choice of $A,B\in\lin{\X}$ we have
\[
\op{vec}(I_\X)^* (A \otimes B) \op{vec}(I_\X)
= \tr(AB^\t).
\]

\item[4.]
\label{item:choi}
Let $A\in\lin{\X,\Y\otimes\Z}$ and suppose $\Phi:\lin{\X}\to\lin{\Y}$ is given
by $\Phi(X) = \tr_\Z(AXA^*)$ for all $X\in\lin{\X}$.
Then
\[
J(\Phi)= \tr_{\Z} (\op{vec}(A)\op{vec}(A)^*).
\]

\end{mylist}
\end{prop}

For any non-empty set $\C\subseteq\herm{\X}$ of Hermitian operators,
the {\it polar} of $\C$ is defined as
\[
\C^{\circ} = \left\{A\in\herm{\X}\,:\,\ip{B}{A}\leq 1 \;
\text{for all $B\in \C$} \right\},
\]
and the {\it support} and {\it gauge} functions of $\C$ are defined as
follows:
\begin{align*}
s(X \mid \C) & = \sup\{\ip{X}{Y}\,:\,Y\in \C\},\\
g(X \mid \C) & = \inf\{\lambda\geq 0\,:\,X \in \lambda\C\}.
\end{align*}
These functions are partial functions in general, but it is typical to
view them as total functions from $\herm{\X}$ to $\mathbb{R} \cup
\{\infty\}$ in the natural way.
For any set $\C$ of positive semidefinite operators, we denote
\[
\downarrow\!\C = \{X\,:\,0\leq X \leq Y\;
\text{for some}\;Y\in\C\}.
\]

\begin{prop} \label{prop:polar}
  Let $\X$ be a complex Euclidean space and let $\C$ and $\D$ be
  non-empty subsets of $\herm{\X}$.
  Then the following facts hold:
  \begin{enumerate}
  \item
    \label{item:polar-contain}
    If $\C\subseteq\D$ then $\D^\circ\subseteq\C^\circ$.
  \item
    \label{item:polar-pos}
    If $-X\in\C$ for each $X\in\pos{\X}$ then
    $\C^\circ\subseteq\pos{\X}$.
  \item
    \label{item:polar-polar}
    If $\C$ is closed, convex, and contains the origin, then the same is
    true of $\C^\circ$.
    In this case we have $\C^{\circ\circ}=\C$,
    \[
    s(\cdot \mid \mathcal{C}) = g(\cdot \mid \mathcal{C}^{\circ}),
    \quad\text{and}\quad
    s(\cdot \mid \mathcal{C}^{\circ}) = g(\cdot \mid \mathcal{C}).
    \]
  \end{enumerate}
\end{prop}

\noindent
The first two items in the above proposition are elementary, and a
proof of the third may be found in Rockafellar~\cite{Rockafellar70}.

%. . . . . . . . . . . . . . . . . . . . . . . . . . . . . . . . . . . . . . .%

\section{Quantum Strategies}
\label{sec:strategies}

\noindent
In this section we define the notions of a quantum {\it strategy} and
the {\it Choi-Jamio{\l}kowski representation} of quantum strategy.
The remainder of the paper is concerned with the study of these objects
and their interactions.

\subsubsection*{Definition of quantum strategies}

We begin with our definition for quantum strategies, which we will
simply call {\it strategies} given that the focus of the paper is on the
quantum setting.

\begin{definition} \label{def:strategy}
  Let $n\geq 1$ and let $\X_1,\ldots,\X_n$ and $\Y_1,\ldots,\Y_n$ be
  complex Euclidean spaces.
  An {\it $n$-turn non-measuring strategy} having input spaces
  $\X_1,\ldots,\X_n$ and output spaces $\Y_1,\ldots,\Y_n$ consists of:
  \begin{mylist}{5mm}
  \item[1.] complex Euclidean spaces $\Z_1,\ldots,\Z_n$, which
    will be called {\it memory spaces}, and
  \item[2.]
    an $n$-tuple of admissible mappings $(\Phi_1,\ldots,\Phi_n)$
    having the form
    \begin{align*}
      \Phi_1 & : \lin{\X_1} \rightarrow \lin{\Y_1\otimes\Z_1}\\
      \Phi_k & : \lin{\X_k\otimes\Z_{k-1}} \rightarrow
      \lin{\Y_k\otimes\Z_k} \quad (2 \leq k \leq n).
    \end{align*}
  \end{mylist}
  An {\it $n$-turn measuring strategy} consists of items 1 and 2
  above, as well as:
  \begin{mylist}{5mm}
  \item[3.]
    a measurement $\left\{P_a\,:\,a\in\Sigma\right\}$ on the last
    memory space $\Z_n$.
  \end{mylist}
  We will use the term {\it $n$-turn strategy} to refer to either a
  measuring or non-measuring $n$-turn strategy.
\end{definition}

\begin{figure*}[t]
  \begin{center}
    \setlength{\unitlength}{0.000245in}
    \begin{picture}(23734,2939)(-600,1700)
      \scriptsize
      \texture{8101010 10000000 444444 44000000 11101 11000000 444444 44000000
	101010 10000000 444444 44000000 10101 1000000 444444 44000000
	101010 10000000 444444 44000000 11101 11000000 444444 44000000
	101010 10000000 444444 44000000 10101 1000000 444444 44000000 }

      \shade\path(18612,4212)(19812,4212)(19812,3012)(18612,3012)(18612,4212)
      \shade\path(1812,4212)(3012,4212)(3012,3012)(1812,3012)(1812,4212)
      \shade\path(5412,4212)(6612,4212)(6612,3012)(5412,3012)(5412,4212)
      \shade\path(9012,4212)(10212,4212)(10212,3012)(9012,3012)(9012,4212)

      \put(2412,3612){\makebox(0,0){$\Phi_1$}}
      \put(6012,3612){\makebox(0,0){$\Phi_2$}}
      \put(9612,3612){\makebox(0,0){$\Phi_3$}}
      \put(19212,3612){\makebox(0,0){$\Phi_n$}}

%      \thicklines
      \path(16512,3612)(18312,3612)
      \blacken\path(18072.000,3552.000)(18312.000,3612.000)(18072.000,3672.000)
      (18072.000,3552.000)

      \path(17712,1512)(18912,2712)
      \blacken\path(18784.721,2499.868)(18912.000,2712.000)(18699.868,2584.721)
      (18784.721,2499.868)

      \path(19512,2712)(20712,1512)
      \blacken\path(20499.868,1639.279)(20712.000,1512.000)(20584.721,1724.132)
      (20499.868,1639.279)

      \path(912,1512)(2112,2712)
      \blacken\path(1984.721,2499.868)(2112.000,2712.000)(1899.868,2584.721)
      (1984.721,2499.868)

      \path(2712,2712)(3912,1512)
      \blacken\path(3699.868,1639.279)(3912.000,1512.000)(3784.721,1724.132)
      (3699.868,1639.279)

      \path(4512,1512)(5712,2712)
      \blacken\path(5584.721,2499.868)(5712.000,2712.000)(5499.868,2584.721)
      (5584.721,2499.868)

      \path(6312,2712)(7512,1512)
      \blacken\path(7299.868,1639.279)(7512.000,1512.000)(7384.721,1724.132)
      (7299.868,1639.279)

      \path(8112,1512)(9312,2712)
      \blacken\path(9184.721,2499.868)(9312.000,2712.000)(9099.868,2584.721)
      (9184.721,2499.868)

      \path(9912,2712)(11112,1512)
      \blacken\path(10899.868,1639.279)(11112.000,1512.000)(10984.721,1724.132)
      (10899.868,1639.279)

      \path(3312,3612)(5112,3612)
      \blacken\path(4872.000,3552.000)(5112.000,3612.000)(4872.000,3672.000)
      (4872.000,3552.000)

      \path(6912,3612)(8712,3612)
      \blacken\path(8472.000,3552.000)(8712.000,3612.000)(8472.000,3672.000)
      (8472.000,3552.000)

      \path(10512,3612)(12312,3612)
      \blacken\path(12072.000,3552.000)(12312.000,3612.000)(12072.000,3672.000)
      (12072.000,3552.000)

      \path(20112,3612)(21912,3612)
      \blacken\path(21672.000,3552.000)(21912.000,3612.000)(21672.000,3672.000)
      (21672.000,3552.000)

      \put(15012,3612){\blacken\ellipse{100}{100}}
      \put(15012,3612){\ellipse{100}{100}}
      \put(13212,3612){\blacken\ellipse{100}{100}}
      \put(13212,3612){\ellipse{100}{100}}
      \put(14112,3612){\blacken\ellipse{100}{100}}
      \put(14112,3612){\ellipse{100}{100}}
      \put(15912,3612){\blacken\ellipse{100}{100}}
      \put(15912,3612){\ellipse{100}{100}}

      \put(1112,2112){\makebox(0,0)[b]{$\X_1$}}
      \put(4712,2112){\makebox(0,0)[b]{$\X_2$}}
      \put(8312,2112){\makebox(0,0)[b]{$\X_3$}}
      \put(17912,2112){\makebox(0,0)[b]{$\X_n$}}

      \put(3662,2112){\makebox(0,0)[b]{$\Y_1$}}
      \put(7262,2112){\makebox(0,0)[b]{$\Y_2$}}
      \put(10862,2112){\makebox(0,0)[b]{$\Y_3$}}
      \put(20612,2112){\makebox(0,0)[b]{$\Y_n$}}

      \put(4212,4012){\makebox(0,0){$\Z_1$}}
      \put(7862,4012){\makebox(0,0){$\Z_2$}}
      \put(11462,4012){\makebox(0,0){$\Z_3$}}
      \put(17462,4012){\makebox(0,0){$\Z_{n-1}$}}
      \put(21012,4012){\makebox(0,0){$\Z_n$}}
    \end{picture}
  \end{center}
  \caption{An $n$-turn strategy.}
  \label{fig:strategy}
\end{figure*}
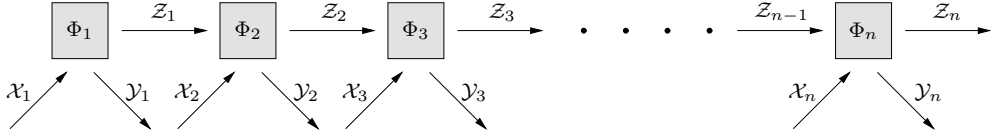

Figure~\ref{fig:strategy} illustrates an $n$-turn non-measuring strategy.

Although there is no restriction on the dimension of the memory spaces
in a quantum strategy, it is established in the proof of
Theorem~\ref{theorem:characterization} that every measuring strategy is
equivalent to one in which
$\dim(\Z_k)\leq\dim(\X_{1\ldots k}\otimes\Y_{1\ldots k})$ for each
$k=1,\ldots,n$.

We also note that our definition of strategies allows the possibility
that any of the input or output spaces is equal to $\complex$, which
corresponds to an empty message.
One can therefore view simple actions such as the preparation of a quantum
state or performing a measurement without producing a quantum output as
special cases of strategies.

When we say that an $n$-turn strategy is described by
{\it linear isometries} $A_1,\ldots,A_n$, it is meant that the
admissible super-operators $\Phi_1,\ldots,\Phi_n$ defining the
strategy are given by $\Phi_k(X) = A_k X A_k^{\ast}$ for $1\leq k\leq n$.
Notice that, when it is convenient, there is no loss of generality in
restricting ones attention to strategies described by linear isometries in
this way.
This is because every admissible super-operator can be expressed as a
mapping $X\mapsto A X A^{\ast}$ for some linear isometry $A$, followed
by the partial trace over some ``garbage'' space that represents a
tensor factor of the space to which $A$ maps.
By including the necessary ``garbage'' spaces as tensor factors of the
memory spaces, and therefore not tracing them out, there can be no change
in the action of the strategy on the input and output spaces.
Along similar lines, there is no loss of generality in assuming
that a given measuring strategy's measurement is projective.

%. . . . . . . . . . . . . . . . . . . . . . . . . . . . . . . . . . . . . . .%

\subsubsection*{Interactions among strategies}

A given $n$-turn strategy expects to interact with something that
provides the inputs corresponding to $\X_1,\ldots,\X_n$
and accepts the strategy's outputs corresponding to $\Y_1,\ldots,\Y_n$.
Let us define an $n$-turn {\it co-strategy} to be the sort of object
that a strategy interfaces with in the most natural way.

\begin{definition}
  Let $n\geq 1$ and let $\X_1,\ldots,\X_n$ and $\Y_1,\ldots,\Y_n$ be
  complex Euclidean spaces.
  The spaces $\X_1,\ldots,\X_n$ are viewed as the input spaces
  of some $n$-turn strategy while $\Y_1,\ldots,\Y_n$ are to be viewed
  as its output spaces.
  An {\it $n$-turn non-measuring co-strategy} to these spaces
  consists of:
  \begin{mylist}{5mm}
  \item[1.] complex Euclidean {\it memory} spaces $\W_0,\ldots,\W_n$,
  \item[2.]
    a density operator $\rho_0 \in \density{\X_1\otimes\W_0}$, and
  \item[3.]
    an $n$-tuple of admissible mappings $(\Psi_1,\ldots,\Psi_n)$
    having the form
    \begin{align*}
      \Psi_k & : \lin{\Y_k\otimes\W_{k-1}} \rightarrow
      \lin{\X_{k+1}\otimes\W_k} \quad (1\leq k\leq n-1)\\
      \Psi_n & : \lin{\Y_n\otimes\W_{n-1}} \rightarrow \lin{\W_n}.
    \end{align*}
  \end{mylist}
  An {\it $n$-turn measuring co-strategy} consists of items 1, 2 and 3
  above, as well as:
  \begin{mylist}{5mm}
  \item[4.]
    a measurement $\left\{Q_b\,:\,b\in\Gamma\right\}$ on the last
    memory space $\W_n$.
  \end{mylist}
  As for strategies, we use the term {\it $n$-turn co-strategy} to
  refer to either a measuring or non-measuring $n$-turn co-strategy.
\end{definition}

\begin{figure*}[t]
  \begin{center}
    \setlength{\unitlength}{0.000245in}
    \begin{picture}(23734,4539)(-600,200)
      \scriptsize

      \shade\path(20412,1212)(21612,1212)(21612,12)(20412,12)(20412,1212)
      \shade\path(18612,4212)(19812,4212)(19812,3012)(18612,3012)(18612,4212)
      \shade\path(1812,4212)(3012,4212)(3012,3012)(1812,3012)(1812,4212)
      \shade\path(3612,1212)(4812,1212)(4812,12)(3612,12)(3612,1212)
      \shade\path(5412,4212)(6612,4212)(6612,3012)(5412,3012)(5412,4212)
      \shade\path(7212,1212)(8412,1212)(8412,12)(7212,12)(7212,1212)
      \shade\path(9012,4212)(10212,4212)(10212,3012)(9012,3012)(9012,4212)
      \shade\path(10812,1212)(12012,1212)(12012,12)(10812,12)(10812,1212)

      \shade\path(12,1212)(1212,1212)(1212,12)(12,12)(12,1212)

      % \put(-1012,3612){\makebox(0,0){$\ket{\psi_0}$}}
      \put(612,612){\makebox(0,0){$\rho_0$}}

      \put(4212,612){\makebox(0,0){$\Psi_1$}}
      \put(7812,612){\makebox(0,0){$\Psi_2$}}
      \put(11412,612){\makebox(0,0){$\Psi_3$}}
      \put(21012,612){\makebox(0,0){$\Psi_n$}}
      \put(2412,3612){\makebox(0,0){$\Phi_1$}}
      \put(6012,3612){\makebox(0,0){$\Phi_2$}}
      \put(9612,3612){\makebox(0,0){$\Phi_3$}}
      \put(19212,3612){\makebox(0,0){$\Phi_n$}}

%      \thicklines

      \path(16512,3612)(18312,3612)
      \blacken\path(18072.000,3552.000)(18312.000,3612.000)(18072.000,3672.000)
      (18072.000,3552.000)

      \path(18312,612)(20112,612)
      \blacken\path(19872.000,552.000)(20112.000,612.000)(19872.000,672.000)
      (19872.000,552.000)

      \path(17712,1512)(18912,2712)
      \blacken\path(18784.721,2499.868)(18912.000,2712.000)(18699.868,2584.721)
      (18784.721,2499.868)

      \path(19512,2712)(20712,1512)
      \blacken\path(20499.868,1639.279)(20712.000,1512.000)(20584.721,1724.132)
      (20499.868,1639.279)

      \path(912,1512)(2112,2712)
      \blacken\path(1984.721,2499.868)(2112.000,2712.000)(1899.868,2584.721)
      (1984.721,2499.868)

      \path(2712,2712)(3912,1512)
      \blacken\path(3699.868,1639.279)(3912.000,1512.000)(3784.721,1724.132)
      (3699.868,1639.279)

      \path(4512,1512)(5712,2712)
      \blacken\path(5584.721,2499.868)(5712.000,2712.000)(5499.868,2584.721)
      (5584.721,2499.868)

      \path(6312,2712)(7512,1512)
      \blacken\path(7299.868,1639.279)(7512.000,1512.000)(7384.721,1724.132)
      (7299.868,1639.279)

      \path(8112,1512)(9312,2712)
      \blacken\path(9184.721,2499.868)(9312.000,2712.000)(9099.868,2584.721)
      (9184.721,2499.868)

      \path(9912,2712)(11112,1512)
      \blacken\path(10899.868,1639.279)(11112.000,1512.000)(10984.721,1724.132)
      (10899.868,1639.279)

%      \path(-312,3612)(1512,3612)
%      \blacken\path(1272.000,3552.000)(1512.000,3612.000)(1272.000,3672.000)
%      (1272.000,3552.000)

      \path(3312,3612)(5112,3612)
      \blacken\path(4872.000,3552.000)(5112.000,3612.000)(4872.000,3672.000)
      (4872.000,3552.000)

      \path(6912,3612)(8712,3612)
      \blacken\path(8472.000,3552.000)(8712.000,3612.000)(8472.000,3672.000)
      (8472.000,3552.000)

      \path(1512,612)(3312,612)
      \blacken\path(3072.000,552.000)(3312.000,612.000)(3072.000,672.000)
      (3072.000,552.000)

      \path(5112,612)(6912,612)
      \blacken\path(6672.000,552.000)(6912.000,612.000)(6672.000,672.000)
      (6672.000,552.000)

      \path(8712,612)(10512,612)
      \blacken\path(10272.000,552.000)(10512.000,612.000)(10272.000,672.000)
      (10272.000,552.000)

      \path(10512,3612)(12312,3612)
      \blacken\path(12072.000,3552.000)(12312.000,3612.000)(12072.000,3672.000)
      (12072.000,3552.000)

      \path(12312,612)(14112,612)
      \blacken\path(13872.000,552.000)(14112.000,612.000)(13872.000,672.000)
      (13872.000,552.000)

      \path(11712,1512)(12912,2712)
      \blacken\path(12784.721,2499.868)(12912.000,2712.000)(12699.868,2584.721)
      (12784.721,2499.868)

      \path(20112,3612)(21912,3612)
      \blacken\path(21672.000,3552.000)(21912.000,3612.000)(21672.000,3672.000)
      (21672.000,3552.000)

      \path(21912,612)(23712,612)
      \blacken\path(23472.000,552.000)(23712.000,612.000)(23472.000,672.000)
      (23472.000,552.000)

      \put(16812,612){\blacken\ellipse{100}{100}}
      \put(16812,612){\ellipse{100}{100}}
      \put(15012,3612){\blacken\ellipse{100}{100}}
      \put(15012,3612){\ellipse{100}{100}}
      \put(13212,3612){\blacken\ellipse{100}{100}}
      \put(13212,3612){\ellipse{100}{100}}
      \put(14112,3612){\blacken\ellipse{100}{100}}
      \put(14112,3612){\ellipse{100}{100}}
      \put(15912,3612){\blacken\ellipse{100}{100}}
      \put(15912,3612){\ellipse{100}{100}}
      \put(15012,612){\blacken\ellipse{100}{100}}
      \put(15012,612){\ellipse{100}{100}}
      \put(15912,612){\blacken\ellipse{100}{100}}
      \put(15912,612){\ellipse{100}{100}}
      \put(17712,612){\blacken\ellipse{100}{100}}
      \put(17712,612){\ellipse{100}{100}}

      \put(1112,2112){\makebox(0,0)[b]{$\X_1$}}
      \put(4712,2112){\makebox(0,0)[b]{$\X_2$}}
      \put(8312,2112){\makebox(0,0)[b]{$\X_3$}}
      \put(11962,2112){\makebox(0,0)[b]{$\X_4$}}
      \put(17912,2112){\makebox(0,0)[b]{$\X_n$}}

      \put(3662,2112){\makebox(0,0)[b]{$\Y_1$}}
      \put(7262,2112){\makebox(0,0)[b]{$\Y_2$}}
      \put(10862,2112){\makebox(0,0)[b]{$\Y_3$}}
      \put(20612,2112){\makebox(0,0)[b]{$\Y_n$}}

 %     \put(612,4012){\makebox(0,0){$\Z_0$}}
      \put(4212,4012){\makebox(0,0){$\Z_1$}}
      \put(7862,4012){\makebox(0,0){$\Z_2$}}
      \put(11462,4012){\makebox(0,0){$\Z_3$}}
      \put(17462,4012){\makebox(0,0){$\Z_{n-1}$}}

      \put(2412,212){\makebox(0,0){$\W_0$}}
      \put(6012,212){\makebox(0,0){$\W_1$}}
      \put(9612,212){\makebox(0,0){$\W_2$}}
      \put(13212,212){\makebox(0,0){$\W_3$}}
      \put(19212,212){\makebox(0,0){$\W_{n-1}$}}
      \put(21012,4012){\makebox(0,0){$\Z_n$}}
      \put(22662,212){\makebox(0,0){$\W_n$}}

    \end{picture}
  \end{center}
  \caption{An interaction between an $n$-turn strategy and co-strategy.}
  \label{fig:interaction}
\end{figure*}
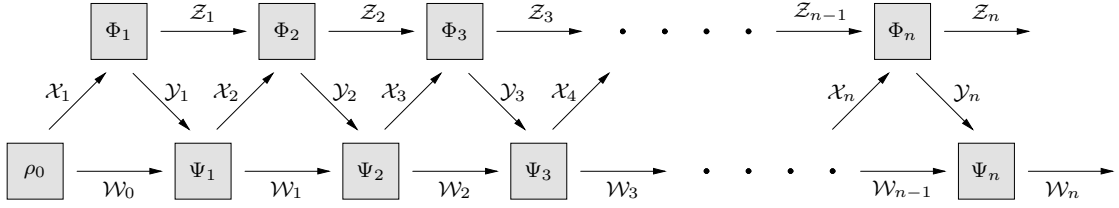

\noindent
Figure~\ref{fig:interaction} represents the interaction between an
$n$-turn strategy and co-strategy.

We have arbitrarily defined strategies and co-strategies in such a way
that the co-strategy sends the first message.
This accounts for the inevitable asymmetry in the definitions.
While it is possible to view any $n$-turn co-strategy as being an
$(n+1)$-turn strategy having input spaces $\complex,\Y_1,\ldots,\Y_n$
and output spaces $\X_1,\ldots,\X_n,\complex$, it will be convenient
for our purposes to view strategies and co-strategies as being
distinct types of objects.

Similar to strategies, there will be no loss of generality in assuming
that the initial state $\rho_0$ of a co-strategy is pure, that each of
the admissible super-operators $\Psi_1,\ldots,\Psi_n$ takes the form
$\Psi_j(X) = B_j X B_j^{\ast}$ for some linear isometry $B_j$, and,
in the case of measuring co-strategies, that the measurement
$\{Q_b\,:\,b\in\Gamma\}$ is a projective measurement.

An $n$-turn strategy and co-strategy are {\it compatible} if they
agree on the spaces $\X_1,\ldots,\X_n$ and $\Y_1,\ldots,\Y_n$.
By the {\it output} of a compatible strategy and co-strategy, assuming at
least one of them is measuring, we mean the result of the measurement
or measurements performed after the interaction between the strategies
takes place.
In particular, if both the strategy and co-strategy make measurements,
then each output $(a,b)\in\Sigma\times\Gamma$ results with probability
\[
\ip{P_a\otimes Q_b}
{(I_{\lin{\Z_n}}\otimes \Psi_n)\cdots(\Phi_1\otimes I_{\lin{\W_0}})\rho_0}.
\]

%The notion of a strategy interacting with a co-strategy as discussed
%above is sufficiently general to model a very wide range of interactions
%among multiple parties.
%For example, a strategy or co-strategy may be induced by the collective
%actions of multiple players.
%This will be the case later, for instance, when we discuss refereed games.

%. . . . . . . . . . . . . . . . . . . . . . . . . . . . . . . . . . . . . . .%

%\subsubsection*{Choi-Jamio{\l}kowski representations of strategies}
\subsubsection*{A new way to represent strategies}

The definitions of strategies and co-strategies given above are
natural from an operational point of view, in the sense that they
clearly describe the actions of the players that they model.
In some situations, however, representing a strategy (or co-strategy)
in terms of a sequence of admissible super-operators is inconvenient.
We now describe a different way to represent strategies that is based
on the Choi-Jamio{\l}kowski representation of super-operators.

Let us first extend this representation to $n$-turn non-measuring
strategies.
To do this, we associate with the strategy described by
$(\Phi_1,\ldots,\Phi_n)$ a single admissible super-operator
\[
\Xi:\lin{\X_{1\ldots n}} \rightarrow \lin{\Y_{1\ldots n}}.
\]
This is the super-operator that takes a given input
$\xi\in\density{\X_{1\ldots n}}$ and feeds the portions of this state
corresponding to the input spaces $\X_1,\ldots,\X_n$ into
the network pictured in Figure~\ref{fig:strategy}, one piece at a time.
The memory space $\Z_n$ is then traced out, leaving some element
$\Xi(\xi)\in \density{\Y_{1\ldots n}}$.
Such a map is depicted in Figure~\ref{fig:choijam} for the case $n=3$.
The Choi-Jamio{\l}kowski representation of the strategy described by
$(\Phi_1,\ldots,\Phi_n)$ is then simply defined as the Choi-Jamio{\l}kowski
representation $J(\Xi)$ of the super-operator $\Xi$ we have just defined.

An alternate expression for the Choi-Jamio{\l}kowski
representation of a strategy exists in the case that it is described
by linear isometries $(A_1,\ldots,A_n)$.
Specifically, its representation is given by
$\tr_{\Z_n}\left(\op{vec}(A)\op{vec}(A)^{\ast}\right)$
where $A\in\lin{\X_{1\ldots n},\Y_{1\ldots n}\otimes\Z_n}$ is defined by the
product
\begin{equation}\label{eq:isometry1}
A =
\left(I_{\Y_{1\ldots n-1}} \otimes A_n\right)
\left(I_{\Y_{1\ldots n-2}}\otimes A_{n-1}\otimes I_{\X_n}\right)\cdots
(A_1 \otimes I_{\X_{2\ldots n}}).
\end{equation}
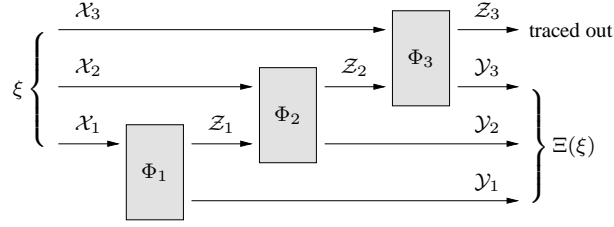
\begin{figure*}[t]
  \begin{center}
  \setlength{\unitlength}{0.000445in}
  \begin{picture}(6999,2697)(0,-10)
    \scriptsize

%    \texture{8101010 10000000 444444 44000000 11101 11000000 444444 44000000
%	101010 10000000 444444 44000000 10101 1000000 444444 44000000
%	101010 10000000 444444 44000000 11101 11000000 444444 44000000
%	101010 10000000 444444 44000000 10101 1000000 444444 44000000 }

    \shade\path(1587,1137)(2262,1137)(2262,12)(1587,12)(1587,1137)
    \shade\path(3162,1812)(3837,1812)(3837,687)(3162,687)(3162,1812)
    \shade\path(4737,2487)(5412,2487)(5412,1362)(4737,1362)(4737,2487)

    \put(1925,575){\makebox(0,0){$\Phi_1$}}
    \put(3500,1250){\makebox(0,0){$\Phi_2$}}
    \put(5075,1925){\makebox(0,0){$\Phi_3$}}

    %\path(6312,2487)(6987,2487)(6987,2037)(6312,2037)(6312,2487)
    %\path(12,462)(687,462)(687,12)(12,12)(12,462)

%    \put(350,237){\makebox(0,0){$\ket{\psi_0}$}}
    \put(6350,2262){\makebox(0,0)[l]{traced out}}
    \put(687,1587){\makebox(0,0)[r]
      {$\xi\left\{\rule[9mm]{0mm}{0mm}\right.$}}
    \put(6312,912){\makebox(0,0)[l]
      {$\left.\rule[9mm]{0mm}{0mm}\right\}\Xi(\xi)$}}

%    \thicklines

    %\path(2262,912)(3162,912)
    %\blacken\path(3042.000,882.000)(3162.000,912.000)
    %  (3042.000,942.000)(3042.000,882.000)
    \path(2362,912)(3062,912)
    \blacken\path(2942.000,882.000)(3062.000,912.000)
      (2942.000,942.000)(2942.000,882.000)

    %\path(687,237)(1587,237)
    %\blacken\path(1467.000,207.000)(1587.000,237.000)
    %  (1467.000,267.000)(1467.000,207.000)
 %   \path(787,237)(1487,237)
 %   \blacken\path(1367.000,207.000)(1487.000,237.000)
 %     (1367.000,267.000)(1367.000,207.000)

    %\path(687,912)(1587,912)
    %\blacken\path(1467.000,882.000)(1587.000,912.000)
    %  (1467.000,942.000)(1467.000,882.000)
    \path(787,912)(1487,912)
    \blacken\path(1367.000,882.000)(1487.000,912.000)
      (1367.000,942.000)(1367.000,882.000)

    %\path(3837,1587)(4737,1587)
    %\blacken\path(4617.000,1557.000)(4737.000,1587.000)
    %  (4617.000,1617.000)(4617.000,1557.000)
    \path(3937,1587)(4637,1587)
    \blacken\path(4517.000,1557.000)(4637.000,1587.000)
      (4517.000,1617.000)(4517.000,1557.000)

    %\path(5412,2262)(6312,2262)
    %\blacken\path(6192.000,2232.000)(6312.000,2262.000)
    %  (6192.000,2292.000)(6192.000,2232.000)
    \path(5512,2262)(6212,2262)
    \blacken\path(6092.000,2232.000)(6212.000,2262.000)
      (6092.000,2292.000)(6092.000,2232.000)

    %\path(5412,1587)(6312,1587)
    %\blacken\path(6192.000,1557.000)(6312.000,1587.000)
    %  (6192.000,1617.000)(6192.000,1557.000)
    \path(5512,1587)(6212,1587)
    \blacken\path(6092.000,1557.000)(6212.000,1587.000)
      (6092.000,1617.000)(6092.000,1557.000)

    %\path(2262,237)(6312,237)
    %\blacken\path(6192.000,207.000)(6312.000,237.000)
    %  (6192.000,267.000)(6192.000,207.000)
    \path(2362,237)(6212,237)
    \blacken\path(6092.000,207.000)(6212.000,237.000)
      (6092.000,267.000)(6092.000,207.000)

    %\path(687,1587)(3162,1587)
    %\blacken\path(3042.000,1557.000)(3162.000,1587.000)
    %  (3042.000,1617.000)(3042.000,1557.000)
    \path(787,1587)(3062,1587)
    \blacken\path(2942.000,1557.000)(3062.000,1587.000)
      (2942.000,1617.000)(2942.000,1557.000)

    %\path(3837,912)(6312,912)
    %\blacken\path(6192.000,882.000)(6312.000,912.000)
    %  (6192.000,942.000)(6192.000,882.000)
    \path(3937,912)(6212,912)
    \blacken\path(6092.000,882.000)(6212.000,912.000)
      (6092.000,942.000)(6092.000,882.000)

    %\path(687,2262)(4737,2262)
    %\blacken\path(4617.000,2232.000)(4737.000,2262.000)
    %  (4617.000,2292.000)(4617.000,2232.000)
    \path(787,2262)(4637,2262)
    \blacken\path(4517.000,2232.000)(4637.000,2262.000)
      (4517.000,2292.000)(4517.000,2232.000)

    \put(5862,2487){\makebox(0,0){$\Z_3$}}
    \put(5862,1812){\makebox(0,0){$\Y_3$}}
    \put(5862,1137){\makebox(0,0){$\Y_2$}}
    \put(5862,462){\makebox(0,0){$\Y_1$}}
    \put(1137,2487){\makebox(0,0){$\X_3$}}
    \put(1137,1812){\makebox(0,0){$\X_2$}}
    \put(1137,1137){\makebox(0,0){$\X_1$}}
%    \put(1137,462){\makebox(0,0){$\Z_0$}}
    \put(4287,1812){\makebox(0,0){$\Z_2$}}
    \put(2712,1137){\makebox(0,0){$\Z_1$}}
  \end{picture}
  \end{center}
  \caption{The super-operator $\Xi$ associated with a 3-turn strategy.}
  \label{fig:choijam}
\end{figure*}

Next we consider measuring strategies.
Assume that an $n$-turn measuring strategy is given, where the
measurement is described by
\[
\{P_a\,:\,a\in\Sigma\}\subset\pos{\Z_n},
\]
for some finite, non-empty set $\Sigma$ of measurement outcomes.
In this case we first associate with the strategy a collection of
super-operators $\{\Xi_a\,:\,a\in\Sigma\}$, each having the form
\[
\Xi_a : \lin{\X_{1\ldots n}} \rightarrow \lin{\Y_{1\ldots n}}.
\]
The definition of each super-operator $\Xi_a$ is precisely as in the
non-measuring case, except that the partial trace over $\Z_n$ is
replaced by the mapping
\[
X\mapsto \tr_{\Z_n} \left( (P_a\otimes I_{\Y_{1\ldots n}}) X\right).
\]
Notice that
\[
\sum_{a\in\Sigma} \Xi_a = \Xi,
\]
where $\Xi$ is the mapping
defined as in the non-measuring case.
Each super-operators $\Xi_a$ is completely positive but generally is not
trace-preserving.
The Choi-Jamio{\l}kowski representation of the measuring strategy
described by $(\Phi_1,\ldots,\Phi_n)$ and $\{P_a\,:\,a\in\Sigma\}$
is defined as $\{J(\Xi_a)\,:\,a\in\Sigma\}$.

In the situation where a measuring strategy is described by linear
isometries $A_1,\ldots,A_n$ and a measurement
$\{P_a\,:\,a\in\Sigma\}$,
its Choi-Jamio{\l}kowski representation is given by
$\{Q_a\,:\,a\in\Sigma\}$ for
\[
Q_a = \tr_{\Z_n}\left( \op{vec}(B_a)\op{vec}(B_a)^{\ast}\right)
\]
where
\[
B_a =
\left(\sqrt{P_a}\otimes I_{\Y_{1\ldots n}}\right) A
\]
for $A$ as defined above \eqref{eq:isometry1}.

It is not difficult to prove that for given input spaces
$\X_1,\ldots,\X_n$ and output spaces $\Y_1,\ldots,\Y_n$,
a collection
$\{Q_a\,:\,a\in\Sigma\}$
of operators
is the Choi-Jamio{\l}kowski representation of some $n$-turn measuring
strategy if and only if $Q = \sum_{a\in\Sigma} Q_a$ is the
representation of an $n$-turn non-measuring strategy over the same
spaces.

Finally, we define the Choi-Jamio{\l}kowski representation of measuring and
non-measuring co-strategies in precisely the same way as for
strategies, except that for technical reasons the resulting operators are
transposed with respect to the standard basis.
(This essentially allows us to eliminate one transposition from almost every
subsequent equation in this paper involving representations of co-strategies.)
Specifically, a given $n$-turn co-strategy is viewed as an $(n+1)$-turn
strategy by including an empty first and last message as discussed
previously.
This strategy's Choi-Jamio{\l}kowski representation
is defined as above.
The operators comprising this strategy representation are then
transposed with respect to the standard basis to obtain the
Choi-Jamio{\l}kowski representation of the co-strategy.

As we work almost exclusively with the Choi-Jamio{\l}kowski
representation of strategies and co-strategies hereafter, we will
typically use the term {\it representation} rather than
{\it Choi-Jamio{\l}kowski representation} for brevity.

%=============================================================================%

\section{Properties of representations}
\label{sec:properties}

\noindent
The applications of Choi-Jamio{\l}kowski representations of strategies
given in this paper rely upon three key properties of these
representations, stated below as Theorems
\ref{theorem:interaction-output-probability},
\ref{theorem:characterization}, and
\ref{theorem:gauge}.
This section is devoted to establishing these properties.

%. . . . . . . . . . . . . . . . . . . . . . . . . . . . . . . . . . . . . . .%

\subsubsection*{Interaction output probabilities}

The first property provides a simple formula for the probability of a given
output of an interaction between a strategy and a co-strategy.

\begin{theorem}\label{theorem:interaction-output-probability}
  Let $\{Q_a\,:\,a\in\Sigma\}$ be the representation of a strategy and
  let $\{R_b \,:\,b \in\Gamma\}$ be the representation of a compatible
  co-strategy.
  For each pair $(a,b) \in \Sigma \times \Gamma$ of measurement
  outcomes, we have that the output of an interaction between
  the given strategy and co-strategy is $(a,b)$ with probability
  $\ip{Q_a}{R_b}$.
\end{theorem}

\begin{proof}
  Let us fix a strategy and co-strategy whose representations are
  $\{Q_a\,:\,a\in\Sigma\}$ and\linebreak
  $\{R_a\,:\,a\in\Gamma\}$, respectively.
  Without loss of generality, the strategy is described by linear
  isometries $A_1,\ldots,A_n$ and a projective measurement
  $\{\Pi_a\,:\,a\in\Sigma\}$, while the co-strategy is described by
  a pure initial state $u_0$, linear isometries
  $B_1,\ldots,B_n$ and a projective measurement
  $\{\Delta_b\,:\,b\in\Gamma\}$.

  For each output pair $(a,b)\in\Sigma\times\Gamma$, define
  $v_{a,b} \in \Z_n\otimes\W_n$ as follows:
  \[
  v_{a,b} =
  (\Pi_a\otimes\Delta_b)
  (I_{\Z_n}\otimes B_n)
  (A_n \otimes I_{\W_{n-1}})
  \cdots
  (I_{\Z_1}\otimes B_1)
  (A_1\otimes I_{\W_0})u_0.
  \]
  The probability of the outcome $(a,b)$ is $\norm{v_{a,b}}^2$
  for each pair $(a,b)$.

  Now, making use of the $\op{vec}$ mapping, we may express $v_{a,b}$
  in a different way:
  \[
  v_{a,b} =
  \left(
  \op{vec}\left(I_{\Y_{1\ldots n}\otimes\X_{1\ldots n}}\right)^{\ast}
  \otimes I_{\Z_n \otimes\W_n}\right)(x_a\otimes y_b),
  \]
  where
  \begin{align*}
    x_a & =
    \left(
      I_{\X_{1\ldots n}\otimes\Y_{1\ldots n}\otimes\Z_n}
      \otimes
      \op{vec}(I_{\Z_{1\ldots n-1}})^{\ast}
    \right)
    (\op{vec}(A_1) \otimes \cdots \otimes
    \op{vec}(A_{n-1})\otimes
    \op{vec}((\Pi_a\otimes I_{\Y_n})
    A_n)),\\
    y_b & =
    \left(
      I_{\X_{1\ldots n}\otimes\Y_{1\ldots n}\otimes\W_n}
      \otimes
      \op{vec}(I_{\W_{0\ldots n-1}})^{\ast}
    \right)
    (u_0\otimes \op{vec}(B_1) \otimes \cdots \otimes
    \op{vec}(B_{n-1})\otimes
    \op{vec}(\Delta_b B_n)).
  \end{align*}
  The probability of outcome $(a,b)$ is therefore
  \begin{align*}
    \norm{v_{a,b}}^2
    & = \tr v_{a,b} v_{a,b}^{\ast}\\
    & =
    \op{vec}(I_{\Y_{1\ldots n}\otimes\X_{1\ldots n}})^{\ast}
    \left[\rule{0mm}{3.6mm}\!
      \left(\tr_{\Z_n} x_a x_a^{\ast}\right)
      \otimes
      \left(\tr_{\W_n} y_b y_b^{\ast}\right)
      \right]
    \op{vec}(I_{\Y_{1\ldots n}\otimes\X_{1\ldots n}})\\
    & = \tr\left[
      \left(\tr_{\Z_n} x_a x_a^{\ast}\right)
      \left(\tr_{\W_n} y_b y_b^{\ast}\right)^{\t}
      \right] \\
    & = \ip{Q_a}{R_b}
  \end{align*}
  as required.
\end{proof}

%. . . . . . . . . . . . . . . . . . . . . . . . . . . . . . . . . . . . . . .%

\subsubsection*{Characterization of strategy representations}

Let us denote by
\[
\st_n(\X_{1\ldots n},\Y_{1\ldots n})
\]
the set of all representations of $n$-turn strategies having input spaces
$\X_1,\ldots,\X_n$ and output spaces $\Y_1,\ldots,\Y_n$.
We may abbreviate this set as $\st_n$ or $\st$ whenever
the spaces or number of turns is clear from the context.
Similarly, we let
\[
\cst_n(\X_{1\ldots n},\Y_{1\ldots n})
\]
denote the set of all
representations of co-strategies for the same spaces.
It will be convenient to define
$\st_0(\complex,\complex) = \cst_0(\complex,\complex) = \{1\}$.

The second property of strategy representations that we prove provides
a characterization of $\st_n(\X_{1\ldots n},\Y_{1\ldots n})$ in terms of linear
constraints on $\pos{\Y_{1\ldots n}\otimes \X_{1\ldots n}}$.
%These linear constraints correspond to the fact that
%(i) each output must be generated before subsequent inputs are
%accessible to the strategy, and
%(ii) the super-operator induced by a strategy must preserve trace.

\begin{theorem} \label{theorem:characterization}
  Let $n\geq 1$, let $\X_1,\ldots,\X_n$ and $\Y_1,\ldots,\Y_n$ be
  complex Euclidean spaces, and let
  $Q\in\pos{\Y_{1\ldots n}\otimes\X_{1\ldots n}}$.
  Then
  \[
  Q \in \st_n(\X_{1\ldots n},\Y_{1\ldots n})
  \]
  if and only if
  \[
  \tr_{\Y_n}(Q) = R \otimes I_{\X_n}
  \]
  for $R \in \st_{n-1}(\X_{1\ldots n-1},\Y_{1\ldots n-1})$.
\end{theorem}

\begin{proof}
First assume that $Q\in\st_n(\X_{1\ldots n},\Y_{1\ldots n})$, which
implies that there exist memory spaces $\Z_1,\ldots,\Z_n$ and admissible
super-operators $\Phi_1, \ldots, \Phi_n$ that comprise a strategy
whose representation is $Q$.
Let $\Xi_n:\lin{\X_{1\ldots n}}\rightarrow\lin{\Y_{1\ldots n}}$
be the super-operator associated with this strategy as described in
Section~\ref{sec:strategies}, and let
\[
\Xi_{n-1}:\lin{\X_{1\ldots n-1}}\rightarrow\lin{\Y_{1\ldots n-1}}
\]
be the super-operator associated with the
$(n-1)$-turn strategy obtained by terminating the strategy described
by $(\Phi_1,\ldots,\Phi_n)$ after $n-1$ turns.
We have
\[
\tr_{\Y_n}(J(\Xi_n)) = J(\tr_{\Y_n} \circ\, \Xi_n)
= J(\Xi_{n-1}\circ \tr_{\X_n}) \\= J(\Xi_{n-1}) \otimes I_{\X_n},
\]
and so $\tr_{\Y_n}(Q) = R\otimes I_{\X_n}$ for
\[
R = J(\Xi_{n-1}) \in \st_{n-1}(\X_{1\ldots n-1},\Y_{1\ldots n-1})
\]
as required.

Now assume that
$Q\in\pos{\Y_{1\ldots n}\otimes\X_{1\ldots n}}$ satisfies
$\tr_{\Y_n}(Q) = R \otimes I_{\X_n}$
for some choice of $R \in \st_{n-1}(\X_{1\ldots n-1},\Y_{1\ldots n-1})$.
Our goal is to prove that $Q \in \st_n(\X_{1\ldots n},\Y_{1\ldots n})$.
This will be proved by induction on $n$.
In fact, it will be easier to prove a somewhat stronger statement,
which is that the assumptions imply that there exists a strategy whose
representation is $Q$ that (i) is described by linear isometries
$A_1,\ldots,A_n$, and (ii) satisfies $\op{dim}(\Z_n) = \op{rank}(Q)$.

If $n=1$, there is nothing new to prove: it is
well-known that if $Q\in\pos{\Y_1\otimes\X_1}$ satisfies
$\tr_{\Y_1} (Q) = I_{\X_1}$, then $Q = J(\Phi_1)$ for some
admissible super-operator
$\Phi_1 : \lin{\X_1}\rightarrow\lin{\Y_1}$.
Any such super-operator can be expressed as
\[
\Phi_1(X) = \tr_{\Z_1} A_1 X A_1^{\ast}
\]
for $\op{dim}(\Z_1) = \op{rank}(Q)$ and some choice of a linear isometry
$A_1\in\lin{\X_1,\Y_1\otimes\Z_1}$.
The 1-turn strategy we require is therefore the strategy described by $A_1$.

Now assume that $n\geq 2$.
By the induction hypothesis, there exist
spaces $\Z_1,\ldots,\Z_{n-1}$ and linear isometries
$A_1,\ldots,A_{n-1}$ with
\begin{align*}
  A_1 & \in \unitary{\X_1,\Y_1\otimes\Z_1},\\
  A_k & \in \unitary{\X_k\otimes\Z_{k-1},\Y_k\otimes\Z_k}
  \quad(2\leq k\leq n-1),
\end{align*}
such that
\[
R = \tr_{\Z_{n-1}}\left(\op{vec}(A)\op{vec}(A)^{\ast}\right)
\]
for $A\in\lin{\X_{1\ldots n-1},\Y_{1\ldots n-1}\otimes\Z_{n-1}}$ defined as
\[
A =(I_{\Y_{1\ldots n-2}} \otimes A_{n-1}) \cdots
(A_1 \otimes I_{\X_{2\ldots n-1}}).
\]

As required, we let $\Z_n$ be a complex Euclidean space with dimension
equal to the rank of $Q$, and let
$B\in\lin{\X_{1\ldots n},\Y_{1\ldots n}\otimes\Z_n}$
be any operator satisfying
\[
\tr_{\Z_n}\left(\op{vec}(B)\op{vec}(B)^{\ast}\right) = Q.
\]
Such a choice of $B$ must exist given that the dimension of $\Z_n$ is
large enough to admit a purification of $Q$.
Note that
\[
\tr_{\Y_n\otimes\Z_n} (\op{vec}(B) \op{vec}(B)^{\ast}) =
\tr_{\Y_n}(Q) =
R \otimes I_{\X_n}.
\]

Next, let $\V$ be a complex Euclidean space with dimension equal to
that of $\X_n$, and let $V\in\unitary{\X_n,\V}$ be an arbitrary
unitary operator.
We have
\[
\tr_{\V}\left(\op{vec}(V)\op{vec}(V)^{\ast}\right) = I_{\X_n},
\]
and therefore
\[
\tr_{\Z_{n-1}\otimes\V}
\left(\op{vec}(A\otimes V)\op{vec}(A\otimes V)^{\ast}\right)
= R \otimes I_{\X_n}.
\]

At this point we have identified two purifications of
$R\otimes I_{\X_n}$.
We will use the isometric equivalence of purifications to define an
isometry $A_n$ that will complete the proof.
Specifically, because $\Z_{n-1}\otimes\V$ has the minimal dimension
required to admit a purification of $R\otimes I_{\X_n}$, it follows
that there must exist a linear isometry
$U\in\unitary{\Z_{n-1}\otimes\V,\Y_n\otimes\Z_n}$
such that
\[
(I_{\Y_{1\ldots n-1}}\otimes U \otimes I_{\X_{1\ldots n}})
\op{vec}(A\otimes V) = \op{vec}(B).
\]
This equation may equivalently be written
\[
B = (I_{\Y_{1\ldots n-1}} \otimes U)(A \otimes V).
\]
We now define $A_n = U(I_{\Z_{n-1}}\otimes V)$.
This is a linear isometry from $\X_n\otimes\Z_{n-1}$ to
$\Y_n\otimes\Z_n$ that satisfies
\[
B = (I_{\Y_{1\ldots n-1}} \otimes A_n)(A\otimes I_{\X_n}).
\]
This implies that the strategy described by $A_1,\ldots,A_n$ has
representation
\[
\tr_{\Z_n}\left(\op{vec}(B)\op{vec}(B)^{\ast}\right)=Q,
\]
and therefore completes the proof.
\end{proof}

Theorem~\ref{theorem:characterization} is equivalent to the following
statement:
an operator $Q\in\pos{\Y_{1\ldots n}\otimes\X_{1\ldots n}}$ is the
representation of some $n$-turn strategy with input spaces
$\X_1,\ldots,\X_n$ and output spaces $\Y_1,\ldots,\Y_n$ if and only if
there exist operators $Q_1,\ldots,Q_{n-1}$, where
$Q_k\in\pos{\Y_{1\ldots k}\otimes\X_{1\ldots k}}$,
such that the following linear constraints are satisfied:
\begin{align*}
\tr_{\Y_{k\ldots n}}(Q) & = Q_{k-1} \otimes I_{\X_{k\ldots n}}
\quad(2\leq k\leq n),\\
\tr_{\Y_{1\ldots n}}(Q) & = I_{\X_{1\ldots n}}.
\end{align*}
Each operator $Q_k$ in of course uniquely determined by the
representation $Q$, and represents the strategy obtained by
terminating any strategy represented by $Q$ after $k$ turns.

Theorem~\ref{theorem:characterization} also gives a characterization
of $n$-turn co-strategies, as stated in the following corollary.

\begin{cor} \label{cor:characterization}
  Let $n\geq 1$, let $\X_1,\ldots,\X_n$ and $\Y_1,\ldots,\Y_n$ be
  complex Euclidean spaces, and let
  $Q\in\pos{\Y_{1\ldots n}\otimes\X_{1\ldots n}}$.
  Then
  \[
  Q \in \cst_n(\X_{1\ldots n},\Y_{1\ldots n})
  \]
  if and only if $Q = R\otimes I_{\Y_n}$
  for $R \in \pos{\Y_{1\ldots n-1}\otimes\X_{1\ldots n}}$ satisfying
  \[
  \tr_{\X_n}(R) \in \cst_{n-1}(\X_{1\ldots n-1},\Y_{1\ldots n-1}).
  \]
\end{cor}

The fact that $\st_n(\X_{1\ldots n},\Y_{1\ldots n})$ and
$\cst_n(\X_{1\ldots n},\Y_{1\ldots n})$ are bounded and characterized by the
positive semidefinite constraint together with finite collections
of linear constraints yields the following corollary.

\begin{cor}
  Let $n\geq 1$, let $\X_1,\ldots,\X_n$ and $\Y_1,\ldots,\Y_n$ be
  complex Euclidean spaces.
  Then the sets $\st_n(\X_{1\ldots n},\Y_{1\ldots n})$ and
  $\cst_n(\X_{1\ldots n},\Y_{1\ldots n})$ are compact and convex.
\end{cor}

%
%Theorem~\ref{theorem:characterization} has established that the sets
%$\st_n$ and $\subst_n$ defined earlier in this section have the following
%inductive characterization:
%$\st_0 = \{1\}$, $\subst_0 = [0,1]$, and
%\begin{align*}
%\st_n &= \left\{ Q
%%\in\pos{\Y_{1\ldots n}\otimes\X_{1\ldots n}}
%\,:\,
%\tr_{\Y_n}(Q) = R\otimes I_{\X_n},\;R\in\st_{n-1}\right\}, \\
%\subst_n &= \left\{ Q
%%\in\pos{\Y_{1\ldots n}\otimes\X_{1\ldots n}}
%\,:\,
%\tr_{\Y_n}(Q) \leq R\otimes I_{\X_n},\;R\in\subst_{n-1}\right\}
%\end{align*}
%for all $n\geq 1$.
%It also follows from that theorem that the sets $\cst_n$ and $\csubst_n$ are
%given by $\cst_0 = \{1\}$, $\csubst_0 = [0,1]$, and
%\begin{align*}
%\cst_n &= \left\{ Q
%%\in\pos{\Y_{1\ldots n}\otimes\X_{1\ldots n}}
%\,:\,
%Q = R\otimes I_{\Y_n},\;\tr_{\X_n}(R) \in \cst_{n-1}\right\}, \\
%\csubst_n &= \left\{ Q
%%\in\pos{\Y_{1\ldots n}\otimes\X_{1\ldots n}}
%\,:\,
%Q \leq R\otimes I_{\Y_n},\;\tr_{\X_n}(R) \in \csubst_{n-1}\right\}
%\end{align*}
%for $n\geq 1$.

Just as $\st_n(\X_{1\ldots n},\Y_{1\ldots n})$ consists of all representations
of non-measuring strategies, the set $\subst_n(\X_{1\ldots n},\Y_{1\ldots n})$
consists of all elements of representations of measuring strategies.
In other words, for any $n$-turn measuring strategy representation
$\{Q_a \,:\, a \in \Sigma\}$, it holds that
$Q_a\in\subst_n(\X_{1\ldots n},\Y_{1\ldots n})$ for each $a\in\Sigma$.
Moreover, for each operator $Q$ there exists an $n$-turn measuring strategy
$\{Q_a\,:\,a\in\Sigma\}$ of which $Q$ is an element if and only if
$Q \in \subst_n(\X_{1\ldots n},\Y_{1\ldots n})$.
The set $\csubst_n(\X_{1\ldots n},\Y_{1\ldots n})$ has similar analogous
properties.

%. . . . . . . . . . . . . . . . . . . . . . . . . . . . . . . . . . . . . . .%

\subsubsection*{Maximum output probabilities}

The final property of strategy representations that we will prove
concerns the maximum probability with which some interacting
co-strategy can force a given measuring strategy to output a given
outcome.

%Theorem~\ref{theorem:gauge} provides a formula that enables the use of
%efficient optimization algorithms for computing the maximum output
%probability of a given strategy.

\begin{theorem} \label{theorem:gauge}
  Let $n\geq 1$, let $\X_1,\ldots,\X_n$ and $\Y_1,\ldots,\Y_n$ be
  complex Euclidean spaces, and let $\{Q_a : a \in \Sigma\}$
  represent an $n$-turn measuring strategy with input
  spaces $\X_1,\ldots,\X_n$ and output spaces $\Y_1,\ldots,\Y_n$.
  Then for each $a\in\Sigma$, the maximum probability
  with which this strategy can be forced to output $a$, maximized over
  all choices of compatible co-strategies, is given by
  \[
  \min\{p\in[0,1]\,:\, Q_a \leq p R\;\;
  \text{for some}\;R\in\st_n(\X_{1\ldots n},\Y_{1\ldots n})\}.
  \]
  An analogous result holds when $\{Q_a\,:\,a\in\Sigma\}$ is a measuring
  co-strategy.
\end{theorem}

\noindent
The remainder of the present section is devoted to a proof of this
theorem.

\begin{lemma} \label{lemma:polar-technical}
  Let $\V$ and $\W$ be complex Euclidean spaces and let
  $\D \subseteq \herm{\V}$ be any closed, convex set that contains the
  origin.
  Then for
  \[
  \C = \left\{ X\in \herm{\V\otimes\W}\,:\, X \leq Y\otimes I_{\W}\;
  \text{\rm for some}\;Y\in\D\right\}
  \]
  we have
  \[
  \C^{\circ} = \left\{ Q \in \pos{\V\otimes\W}\,:\,\tr_{\W}(Q) \in \D^{\circ}
  \right\}.
  \]
\end{lemma}

\begin{proof}
  The assumption $0\in\D$ implies that $-R \in \C$ for every
  $R\in\pos{\V\otimes\W}$, and therefore
  $\C^{\circ} \subseteq \pos{\V\otimes\W}$.
  Consider any choice of $Q\in\pos{\V\otimes\W}$, and note that
  \[
  \ip{\tr_{\W}(Q)}{Y} = \ip{Q}{Y\otimes I_{\W}}
  \]
  for every choice of $Y\in\herm{\V}$.

  If it is the case that $Q\in\C^{\circ}$ then we have
  $\ip{Q}{Y\otimes I_{\W}} \leq 1$ for all $Y\in\D$, and therefore
  $\tr_{\W}(Q) \in \D^{\circ}$.
  On the other hand, if $\tr_{\W}(Q) \in \D^{\circ}$ then
  $\ip{Q}{Y\otimes I_{\W}} \leq 1$
  for all $Y\in\D$.
  It follows from the fact that $Q$ is positive semidefinite
  that $\ip{Q}{X}\leq 1$ for all $X\leq Y\otimes I_{\W}$, and
  therefore $Q\in\C^\circ$.
\end{proof}

\begin{lemma} \label{lemma:switch-polar}
  Let $\V$ be a complex Euclidean space, let $\A,\B\subset\pos{\V}$
  be non-empty closed and convex sets, and suppose
  \[
  (\downarrow\!\A)^{\circ} = \left\{X\in\herm{\V}\,:\,X\leq Q \;\;
    \text{for some}\; Q\in\B\right\}.
  \]
  Then
  \[
  (\downarrow\!\B)^{\circ} = \left\{Y\in\herm{\V}\,:\,Y\leq R \;\;
    \text{for some}\; R\in\A\right\}.
  \]
\end{lemma}

\begin{proof}
Let
\[
\C = \left\{Y\in\herm{\V}\,:\,Y\leq R \;
  \text{for some}\; R\in\A\right\}.
\]
As $-P\in\C$ for every $P\in\pos{\V}$, it follows that
$\C^{\circ}\subseteq\pos{\V}$.
Clearly $\downarrow\!\A \subseteq\C$, and therefore
$\C^{\circ}\subseteq(\downarrow\!\A)^{\circ}$.
Thus,
\[
\C^{\circ}\subseteq (\downarrow\!\A)^{\circ}\cap\pos{\V} =
\:\downarrow\!\B.
\]
On the other hand, we have that every $Q\in\:\downarrow\!\B$ is
contained in $(\downarrow\!\A)^{\circ}$, implying that
$\ip{Q}{R}\leq 1$ for all $R\in\A$.
As $Q\geq 0$, this implies that $\ip{Q}{X}\leq 1$ for $X\leq R$.
Consequently, $Q\in \C^{\circ}$.
Thus $\downarrow\!\B = \C^{\circ}$, and so $(\downarrow\!\B)^{\circ} =
\C$ as required.
\end{proof}

\begin{lemma}\label{lemma:polar}
Let $n\geq 1$ and let $\X_1,\ldots,\X_n$ and $\Y_1,\ldots,\Y_n$ be
complex Euclidean spaces.
Then for all $X\in\herm{\Y_{1\ldots n}\otimes\X_{1\ldots n}}$ we have
\begin{mylist}{5mm}
\item[1.]
$X\in (\subst_n(\X_{1\ldots n},\Y_{1\ldots n}))^\circ$ if and only if
$X\leq Q$ for some choice of $Q\in\cst_n(\X_{1\ldots n},\Y_{1\ldots n})$.

\item[2.]
$X\in(\csubst_n(\X_{1\ldots n},\Y_{1\ldots n}))^\circ$ if and only if
$X\leq Q$ for some choice of $Q\in\st_n(\X_{1\ldots n},\Y_{1\ldots n})$.
\end{mylist}
\end{lemma}

\begin{proof}
  The proof is by induction on $n$.
  As $\subst_0 = \csubst_0 =  [0,1]$
  and $(\subst_0)^{\circ} = (\csubst_0)^{\circ} =  (-\infty,1]$, the
  lemma holds for the case $n = 0$.

  Now suppose that $n\geq 1$.
  The two items in the statement of the lemma are equivalent by
  Lemma~\ref{lemma:switch-polar}, so it will suffice to prove the
  first.

  Define $\E \subset \herm{\Y_{1\ldots n-1}\otimes \X_{1\ldots n}}$ as
  \[
  \E = \left\{ Y \,:\, Y \leq P\otimes I_{\X_n}\;\text{for some}\; P \in
    \subst_{n-1} \right\}.
  \]
  By Lemma~\ref{lemma:polar-technical} we have
  \[
  \E^{\circ} = \left\{
    Q\in\pos{\Y_{1\ldots n-1}\otimes\X_{1\ldots n}}\,:\,
    \tr_{\X_n}(Q) \in (\subst_{n-1})^{\circ}\right\}.
  \]
  Also define $\F\subset \herm{\Y_{1\ldots n}\otimes \X_{1\ldots n}}$ as
  \[
  \F = \left\{ Z \,:\, Z \leq Q\otimes I_{\Y_n}\;\text{for some}\; Q \in
    \E^{\circ} \right\}.
  \]
  Again applying Lemma~\ref{lemma:polar-technical}, we obtain
  \[
  \F^{\circ} =  \left\{
    R\in\pos{\Y_{1\ldots n}\otimes\X_{1\ldots n}}\,:\,
    \tr_{\Y_n}(R) \in \E \right\}.
  \]

  Now, by Theorem~\ref{theorem:characterization} we have
  $\F^{\circ} = \subst_n(\X_{1\ldots n},\Y_{1\ldots n})$, and so
  $\F = (\subst_n(\X_{1\ldots n},\Y_{1\ldots n}))^\circ$.
  By the induction hypothesis we have
  \[
  \E^{\circ} = \left\{
    Q\in\pos{\Y_{1\ldots n-1}\otimes\X_{1\ldots n}}\,:\,
    \tr_{\X_n}(Q) \in \csubst_{n-1}\right\},
  \]
  and therefore
  \[
  \F = \left\{Z \,:\, Z \leq Q\otimes I_{\Y_n}\;\text{for}\;
    \tr_{\X_n}(Q) \in \csubst_{n-1}\right\}.
  \]
  By Corollary~\ref{cor:characterization} we have that
  \[
  \F = \left\{Z \,:\, Z \leq R\;\text{for some}\;
    R \in \csubst_n\right\},
  \]
  which completes the proof.
\end{proof}

\begin{proof}[Proof of Theorem~\ref{theorem:gauge}]
  Let $p_a\in[0,1]$ denote the maximum probability with which
  $\{Q_a\,:\,a\in\Sigma\}$ can be forced to output $a$ in an
  interaction with some compatible co-strategy.
  It follows from Theorem~\ref{theorem:interaction-output-probability} that
  $p_a = s(Q_a \mid \cst)$.
  Using Lemma~\ref{lemma:polar}, along with the fact that $Q_a$ is positive
  semidefinite, we have
  \[
  s(Q_a \mid \cst) = s(Q_a \mid \csubst) = g(Q_a \mid (\csubst)^{\circ})\\
  = g(Q_a \mid \subst),
  \]
  which completes the proof.
\end{proof}

%=============================================================================%

\section{Applications}

\subsubsection*{Kitaev's bound for strong coin-flipping}

Quantum coin-flipping protocols aim to solve the following problem:
two parties, Alice and Bob, at separate locations, want to flip a coin but do
not trust one another.
A quantum coin-flipping protocol with bias $\varepsilon$ is an interaction
between two (honest) strategies $A$ and $B$, both having output sets
$\{0,1,\text{abort}\}$, that satisfies two properties:
\begin{enumerate}
\item
The interaction between the honest parties $A$ and $B$ produces the same
outcome $b\in\{0,1\}$ for both players, with probability $1/2$ for each
outcome.
(Neither player outputs ``abort'' when both are honest.)
\item
If one of the parties does not follow the protocol but the other does,
neither of the outcomes $b\in\{0,1\}$ is output by the honest player
with probability greater than $1/2 + \varepsilon$.
\end{enumerate}

\noindent
Protocols that satisfy these conditions are generally referred to as
{\it strong} coin-flipping protocols, because they require that a cheater
cannot bias an honest player's outcome toward either result 0 or 1.
(In contrast, {\it weak} protocols assume that Alice desires outcome 0 and
Bob desires outcome 1, and only require that cheaters cannot bias the
outcome toward their desired outcome.)

Kitaev \cite{Kitaev02} proved that no strong quantum coin-flipping protocol
can have bias smaller than $1/\sqrt{2} - 1/2$, meaning that one cheating party
can always force a given outcome on an honest party with probability at least
$1/{\sqrt{2}}$.
Kitaev did not publish this proof, but it appears in
Refs.~\cite{AmbainisB+04,Roehrig04}.
Here we give a different proof based on the results of the previous section.

Suppose $\{A_0,A_1,A_{\text{abort}}\}$ is the representation of
honest-Alice's strategy and $\{B_0,B_1,B_{\text{abort}}\}$ is the
representation of honest-Bob's co-strategy in some coin-flipping protocol.
These strategies may involve any fixed number of rounds of interaction.
The first condition above implies
\[
\frac{1}{2} = \ip{A_0}{B_0} = \ip{A_1}{B_1}.
\]

Now, fix $b\in\{0,1\}$, and let $p$ be the maximum probability that a
cheating Bob can force honest-Alice to output $b$.
Obviously we have $p\geq 1/2$.
Theorem~\ref{theorem:gauge} implies that there must exist a strategy
$Q$ for Alice such that $A_b \leq p\, Q$.
If a cheating Alice plays this strategy $Q$, then
honest-Bob outputs $b$ with probability
\[
\ip{Q}{B_b} \geq \frac{1}{p} \ip{A_b}{B_b} = \frac{1}{2p}.
\]
Given that
\[
\max\left\{p,\frac{1}{2p}\right\}\geq \frac{1}{\sqrt{2}}
\]
for all $p > 0$, we have that either honest-Alice or honest-Bob can be
convinced to output $b$ with probability at least $1/{\sqrt{2}}$.

This proof makes clear the limitations of strong coin-flipping
protocols: the inability of Bob to force Alice to output $b$ directly
implies that Alice can herself bias the outcome toward $b$.
Weak coin-flipping does not directly face this same limitation.
Currently the best bound known on the bias of weak quantum coin-flipping
protocols, due to Ambainis \cite{Ambainis01}, is that
$\Omega(\log\log (1/\varepsilon))$ rounds of communication
are necessary to achieve a bias of $\varepsilon$.
The best weak quantum coin-flipping protocol currently known, which is due to
Mochon \cite{Mochon04}, achieves bias approximately 0.192
(which surpasses the barrier $1/\sqrt{2} - 1/2 \approx 0.207$ on strong
quantum coin-flipping).

%. . . . . . . . . . . . . . . . . . . . . . . . . . . . . . . . . . . . . . .%

\subsubsection*{Zero-sum quantum games}

Next, we define quantum refereed games.
It will be noted that von Neumann's Min-Max Theorem for zero-sum quantum
refereed games follows from the facts we have proved about
representations of strategies together with well-known generalizations of
the classical Min-Max Theorem.
Although it is completely expected that the Min-Max Theorem should hold for
quantum games, it has not been previously noted in the general case with which
we are concerned.
(Lee and Johnson \cite{LeeJ03} proved this fact in the one-round case.)
This discussion will also be helpful for the application to
interactive proof systems with competing provers that follows.

Let us first define specifically what is meant by a zero-sum quantum
refereed game.
Such a game is played between two players, Alice and Bob, and is arbitrated by
a referee.
The referee's output after interacting with Alice and Bob for some
fixed number of rounds determines their pay-offs.

\begin{definition} \label{def:referee}
  An {\it $n$-turn referee} is an $n$-turn measuring co-strategy
  $\{R_a\,:\,a\in\Sigma\}$
  whose input spaces $\X_1,\ldots,\X_n$ and output spaces
  $\Y_1,\ldots,\Y_n$ take the form
  \[
  \X_k = \A_k\otimes\B_k\quad\quad\text{and}\quad\quad\Y_k = \C_k\otimes\D_k
  \]
  for complex Euclidean spaces $\A_k$, $\B_k$, $\C_k$ and $\D_k$, for
  $1\leq k \leq n$.
  An {\it $n$-turn quantum refereed game} consists of an $n$-turn
  referee along with functions
  \[
  V_A,V_B:\Sigma\rightarrow\mathbb{R}
  \]
  defined on the referee's set of measurement outcomes, representing
  Alice's payoff and Bob's payoff for each output $a\in\Sigma$.
  Such a game is a {\it zero-sum} quantum refereed game
  if $V_A(a) + V_B(a) = 0$ for all $a\in\Sigma$.
\end{definition}

The referee's actions in a quantum refereed game are completely
determined by its representation $\{R_a\,:\,a\in\Sigma\}$.
During each turn, the referee simultaneously sends a message to Alice
and a message Bob, and a response is expected from each player.
The spaces $\A_k$ and $\B_k$ correspond to the messages sent by the
referee during turn number $k$, while $\C_k$ and $\D_k$ correspond to their
responses.
After $n$ turns, the referee produces an output $a\in \Sigma$.

A refereed quantum game does not in itself place any restrictions on
the strategies available to Alice and Bob.
For example, Alice and Bob might utilize a strategy that allows quantum
communication, they might share entanglement but be forbidden from
communicating, or might even be forbidden to share entanglement.
Specific characteristics of a given game, such as its Nash equilibria,
obviously depend on such restrictions in general.

The focus of the remainder of the paper is on the comparatively simple
setting of zero-sum quantum refereed games.
In this case, we assume that Alice and Bob do not communicate or
share entanglement before the interaction takes place.
More precisely, we assume that Alice and Bob play independent strategies
represented by
$A\in\st_n(\A_{1\ldots n},\C_{1\ldots n})$
and $B\in\st_n(\B_{1\ldots n},\D_{1\ldots n})$,
respectively.
The combined actions of Alice and Bob are therefore together described by
the operator $A\otimes B\in\st_n(\X_{1\ldots n},\Y_{1\ldots n})$.

It is a completely natural assumption that Alice and Bob play
independent strategies in a zero-sum quantum refereed game,
given that it cannot simultaneously be to both players' advantage to
communicate directly with one another or to initially share an
entangled state.
This should not be confused with the possibility that entanglement
among the players and referee might exist at various points in the
game, or that the referee might choose to pass information from one
player to the other.
These possibilities are not disallowed when Alice and Bob's joint
strategy is represented by $A\otimes B$.

Now, assume that a zero-sum quantum refereed game is given, and that
Alice and Bob play independent strategies $A$ and $B$ as just discussed.
Let us write
\[
V(a) = V_A(a) = -V_B(a),
\]
and define
\[
R = \sum_{a\in\Sigma}V(a) R_a.
\]
Alice's expected pay-off is then given by
\[
\sum_{a\in\Sigma}V(a) \ip{A\otimes B}{R_a} = \ip{A\otimes B}{R},
\]
while Bob's expected pay-off is $-\ip{A\otimes B}{R}$.

Now, $\ip{A\otimes B}{R}$ is a real-valued bilinear function in $A$
and $B$.
Because the operators $A$ and $B$ are drawn from compact, convex sets
$\st_n(\A_{1\ldots n},\C_{1\ldots n})$ and
$\st_n(\B_{1\ldots n},\D_{1\ldots n})$ respectively, we have that
\[
\max_{A\in\st_n(\A_{1\ldots n},\C_{1\ldots n})}\,
\min_{B\in\st_n(\B_{1\ldots n},\D_{1\ldots n})}
\ip{A\otimes B}{R} 
=
\min_{B\in\st_n(\B_{1\ldots n},\D_{1\ldots n})}\,
\max_{A\in\st_n(\A_{1\ldots n},\C_{1\ldots n})}
\ip{A\otimes B}{R}.
\]
This is the Min-Max Theorem for zero-sum quantum games.

Note that the above expression does not immediately follow from
von~Neumann's original Min-Max Theorem, but follows from
an early generalization due to J.~Ville \cite{Ville38} and several subsequent
generalizations such as the well-known Min-Max Theorem of
Ky~Fan~\cite{Fan53}.
The real number represented by the two sides of this equation is
called the {\it value} of the given game.

%. . . . . . . . . . . . . . . . . . . . . . . . . . . . . . . . . . . . . . .%

\subsubsection*{Quantum interactive proofs with competing provers}

Classical interactive proof systems with competing provers have been
studied by several authors, including Feige, Shamir, and Tennenholts
\cite{FeigeS+90}, Feige and Shamir \cite{FeigeS92}, Feigenbaum,
Koller, and Shor \cite{FeigenbaumK+95}, and Feige and Kilian
\cite{FeigeK97}.
A quantum variant of this model is defined by allowing the verifier to
exchange quantum information with the provers
\cite{GutoskiW05,Gutoski05}.
In both cases these interactive proof systems are generalizations of
single prover interactive proof systems
\cite{Babai85,BabaiM88,GoldwasserM+89,KitaevW00,Watrous03-pspace}.

Interactive proof systems with two competing provers are naturally
modeled by zero-sum refereed games.
To highlight this connection we will refer to the verifier as the
referee and the two provers as Alice and Bob.
The referee is assumed to be computationally bounded while Alice and
Bob are computationally unrestricted.
Alice, Bob, and the referee receive a common input string
$x\in\{0,1\}^{\ast}$, and an interaction follows.
After the interaction takes place, the referee decides that either
Alice wins or Bob wins.

A language or promise problem $L = (L_{\yes},L_{\no})$ is said to have
a classical refereed game if there exists a referee, described by
a polynomial-time randomized computation, such that:
(i) for every input $x\in L_{\yes}$, there is a strategy for Alice
that wins with probability at least $3/4$ against every strategy of
Bob, and (ii) for every input $x\in L_{\no}$, there is a strategy
for Bob that wins with probability at least $3/4$ against every
strategy of Alice.

The class of promise problems having classical refereed games is
denoted $\class{RG}$.
It is known that $\class{RG}$ is equal to $\class{EXP}$.
%Koller and Megiddo \cite{KollerM92} proved that
The work of Koller and Megiddo \cite{KollerM92} implies
$\class{RG} \subseteq \class{EXP}$,
while Feige and Kilian \cite{FeigeK97} proved the reverse containment.

Let us note that zero-sum classical refereed games, and therefore the
class $\class{RG}$, are unaffected by the
assumption that Alice and Bob may play quantum strategies, assuming
the referee remains classical.
This assumes of course that the classical referee is modeled properly
within the setting of quantum information, which requires that any
quantum information that it touches immediately loses coherence.
Equivalently, the referee effectively measures all messages sent to it
by Alice and Bob with respect to the standard basis before any further
processing takes place.
As there also cannot be a mutual advantage to Alice and Bob to
correlate their strategies using shared entanglement, there is no
advantage to Alice or Bob to play a quantum strategy against a
classical referee.
This is not the case in the cooperative setting, because there Alice
and Bob might use a shared entangled state to their
advantage~\cite{CleveH+04}.

Quantum interactive proof systems with competing provers are defined
in a similar way to the classical case, except that the referee's
actions are described by polynomial-time generated quantum circuits
and the referee may exchange quantum information with Alice and Bob.
The complexity class of all promise problems having quantum refereed
games is denoted $\class{QRG}$.
The containment $\class{EXP} \subseteq \class{QRG}$ follows from
$\class{EXP} \subseteq \class{RG}$.
It was previously known that
$\class{QRG}\subseteq \class{NEXP}\cap\class{co-NEXP}$
\cite{Gutoski05}, and we will improve this to
$\class{QRG}\subseteq\class{EXP}$.
This establishes the characterization $\class{QRG} = \class{EXP}$, and
implies that quantum and classical refereed games are equivalent with
respect to their expressive power.

In the refereed game associated with a competing prover quantum
interactive proof system, the referee declares either Alice or
Bob to be the winner.
Specifically, the referee outputs one of two possible values $\{a,b\}$,
with $a$ meaning that Alice wins and $b$ meaning that Bob wins.
By setting $V(a) = 1$ and $V(b) = 0$, we obtain a quantum refereed
game whose value is the maximum probability with which Alice can win.
We will show that this optimal winning probability for Alice can be
efficiently approximated: it is the value of a semidefinite programming
problem whose size is polynomial in the total dimension of the input and
output spaces of the referee.
It follows from the polynomial-time solvability of semidefinite
programming problems \cite{Khachiyan79,GroetschelL+88,NesterovN94}
that $\class{QRG}\subseteq\class{EXP}$.
%(It also follows that the value of an arbitrary quantum zero-sum game
%can be efficiently approximated.)

We will use similar notation to the previous subsection:
for a fixed
input $x$, the referee is represented by operators $\{R_a,R_b\}$,
and assuming $n$ is the number of turns for this referee we let
the input and output spaces to Alice be denoted by
$\A_1,\C_1,\ldots,\A_n,\C_n$ while $\B_1,\D_1,\ldots,\B_n,\D_n$
denote the input and output spaces to Bob.
The assumption that the referee is described by polynomial-time
generated quantum circuits implies that the entries in the matrix
representations of $R_a$ and $R_b$ with respect to the standard basis
can be approximated to very high precision in exponential time.

Now, given any strategy $A\in\st_n(\A_{1\ldots n},\C_{1\ldots n})$ for
Alice, let us define
\begin{align*}
\Omega_a(A) &= \tr_{\C_{1\ldots n}\otimes\A_{1\ldots n}}
\left((A\otimes I_{\D_{1\ldots n}\otimes\B_{1\ldots n}})R_a\right),
\\
\Omega_b(A) &= \tr_{\C_{1\ldots n}\otimes\A_{1\ldots n}}
\left((A\otimes I_{\D_{1\ldots n}\otimes\B_{1\ldots n}})R_b\right).
\end{align*}
The functions $\Omega_a$ and $\Omega_b$ are linear and
extend to uniquely defined super-operators.
Under the assumption that Alice plays the strategy represented
by $A\in\st_n(\A_{1\ldots n},\C_{1\ldots n})$ and Bob plays the strategy
represented by $B\in\st_n(\B_{1\ldots n},\D_{1\ldots n})$, we have that the
referee outputs $a$ with probability
$\ip{A\otimes B}{R_a} = \ip{B}{\Omega_a(A)}$
and outputs $b$ with probability
$\ip{A\otimes B}{R_b} = \ip{B}{\Omega_b(A)}$.
One may think of $\{\Omega_a(A),\Omega_b(A)\}$ as being the
co-strategy that results from ``hard-wiring'' Alice's strategy
represented by $A$ into the referee.

Now, Alice's goal is to minimize the maximum probability with which Bob can
win.
For a given strategy $A$ for Alice, the maximum probability with which
Bob can win is
\[
\max\left\{ \ip{B}{\Omega_b(A)}\,:\,
B\in\st_n(\B_{1\ldots n},\D_{1\ldots n})\right\}
\]
which, by Theorem~\ref{theorem:gauge}, is given by
\[
\min\left\{ p\geq 0:
\Omega_b(A)\leq p\,Q,\:
Q\in\cst_n(\B_{1\ldots n},\D_{1\ldots n})\right\}.
\]
The following optimization problem therefore determines the maximum
probability $p$ for Bob to win, minimized over all strategies for Alice:
\begin{alignat*}{1}
    \text{Minimize:} \quad & p \\
    \text{Subject to:} \quad
    & \Omega_b(A)\leq p\,Q,\\
    & A\in\st_n(\A_{1\ldots n},\C_{1\ldots n}), \\
    & Q\in\cst_n(\B_{1\ldots n},\D_{1\ldots n}).
\end{alignat*}
This optimization problem can be expressed in terms of linear and
semidefinite constraints as follows:
\begin{alignat*}{2}
    \text{Minimize:} \quad & \tr(P_1) \\
    \text{Subject to:} \quad
    & \,\Omega_b(A_n)\leq Q_n,\\[2mm]
    & \tr_{\C_k}(A_k) = A_{k-1} \otimes I_{\A_k} \quad & (2\leq k\leq n),\\
    & \tr_{\C_1}(A_1) = I_{\A_1},\\[2mm]
    & Q_k = P_k \otimes I_{\D_k} \quad & (1\leq k\leq n),\\
    & \tr_{\B_k}(P_k) = Q_{k-1} \quad & (2\leq k\leq n),\\[2mm]
    & A_k \in \pos{\C_{1\ldots k}\otimes\A_{1\ldots k}} \quad & (1\leq k\leq n),
    \\
    & Q_k \in \pos{\D_{1\ldots k}\otimes\B_{1\ldots k}} \quad & (1\leq k\leq n),
    \\
    & P_k \in \pos{\D_{1\ldots k-1}\otimes\B_{1\ldots k}}\quad & (1\leq k\leq
    n).
\end{alignat*}

%=============================================================================%

\subsection*{Acknowledgements}

This research was supported by Canada's NSERC and the Canadian
Institute for Advanced Research (CIAR).

%=============================================================================%

%\bibliographystyle{acm}
%\bibliography{Bibliography}

\end{document}